\documentclass[12pt,preprint]{aastex}

\slugcomment{Submitted, ApJ, 03/08/2010}
\shorttitle{HAT-P-6b TTVs}
\shortauthors{Dittmann et al.}

\begin{document}

\title{Transit Observations of the WASP-10 System}

\author{Jason A. Dittmann, Laird M. Close, Louis J. Scuderi, \& Marita D. Morris}
\affil{Steward Observatory, University of Arizona, Tucson, AZ 85721}

\begin{abstract}
We present here observations of the transit of WASP-10b on 14 October 2009 UT taken from the University of Arizona's 1.55 meter Kuiper telescope on Mt. Bigelow. Conditions were photometric and accuracies of 2.0 mmag RMS were obtained throughout the transit. We have found that the ratio of the planet to host star radii is in agreement with the measurements of Christian et al. (2008) instead of the refinements of Johnson et al. (2009), suggesting that WASP-10b is indeed inflated beyond what is expected from theoretical modeling. We find no evidence for large ($> 20$ s) transit timing variations in WASP-10b's orbit from the ephemeris of Christian et al. (2008) and Johnson et al. (2009).
\end{abstract}
\keywords{planetary systems, stars: individual: WASP-10}

\section{Introduction}
Transiting Extrasolar planets are relatively rare among all known planet discoveries (69 transiting planets out of 429 known planetary systems)\footnote{http://exoplanet.eu}. However, transits are unique in that they allow for the direct measurement of the radius of the transiting planet relative to the host star. Combined with radial velocity data, it is possible to determine the density of the transiting planet as well. Knowledge of the heating of the star can allow models of the bulk properties of these planets to be formulated and then compared to observation (for example, Baraffe et al. 2008; Fortney et al. 2007; Burrows et al 2007).

Since the depth of the transit is directly related to its radius, the size of a transiting planet is one of the easiest parameters to measure. This makes transiting planets especially interesting to discover and study in detail, however this method becomes increasingly difficult when trying to probe to lower mass planets, although the Kepler Space Telescope (Borucki et al. 2008) is expected to detect many planets as small, or smaller, than the Earth. 

While most transiting planets are in nearly circular orbits, there are a small number of planets that maintain eccentric orbits despite being older than the circularization time scale for the system. The most notable of these systems are the transiting Neptunes GJ 436b (Butler et al. 2004)  and HAT-P-11b (Bakos et al. 2010). GJ 436b has a significantly nonzero eccentricity of $0.15 \pm 0.012$ (Deming et al. 2007), despite being older than its circularization time scale (Maness et al. 2007), which has led to speculation about possible additional (possibly resonant) planets in the system gravitationally ``pumping" the eccentricity. Gravitational interaction with a third body would also lead to transit timing variations as the line of nodes precesses, but searches for this effect have found no evidence for transit timing variations in this system (Pont et al. 2009; Ballard et al. 2010). A search in the similar HAT-P-11 exo-Neptune system has yielded no positive indication for large transit timing variations (Dittmann et al. 2009a). 

WASP-10b is a recently discovered 2.96 M$_J$, 1.28 R$_J$ hot Jupiter in orbit around a K5 dwarf star (Christian et al. 2008). Due to the small size of the star ($0.775^{+0.043}_{-0.040} R_\sun$), the transit depth is relatively deep, at 29 mmag (Christian et al. 2008). WASP-10b is in a $3.0927636^{+0.0000094}_{-0.000021}$ day, $0.059^{+0.014}_{-0.004}$ eccentricity orbit (Christian et al. 2008).  Christian et al. (2008) note that the existence of a significantly nonzero eccentricity is surprising given the system's age (600 Myr - 1Gyr), and that studies into tidal dissipation by Jackson et al. (2008) suggest that WASP-10b's orbit should currently be circularized.  

Johnson et al. (2009) recently performed high quality follow up observations of a transit of WASP-10b using the Orthogonal Parallel Transfer Imaging Camera (OPTIC) on the University of Hawaii 2.2 m telescope on Mauna Kea. Using OPTIC's unique charge transfer\footnote{To our knowledge, this was the first time this technique was used for an extrasolar planet transit.}, they were able to significantly improve the precision in the measurement of various parameters of the system, including the transit time to within 7 seconds uncertainty and the planetary radius to within $1.8\%$ uncertainty. Johnson et al. (2009) note that their error in the planetary radius is dominated by the error in the radius of the star. Furthermore, their value of $1.080 \pm 0.020$ R$_J$ represents a surprising $16\%$ downward change ($5\sigma$ lower than the value of Christian et al. (2008)) in the radius of the planet, corresponding to the value expected by models from Fortney et al. (2007). An attempt to confirm a smaller radius for WASP-10 and investigate the possiblity of transit timing variations in the system due to a second planet acting as an eccentricity pump were the primary motivations for this paper.

\section{Observations \& Reductions}

We observed the transit of WASP-10b on 14 October, 2009 UT using the University of Arizona's 61 inch (1.55 meter) Kuiper telescope on Mt.\ Bigelow, Arizona with the Mont4k CCD, binned 3x3 to 0.43$^{\prime\prime}$/pixel. Our observations were taken with an Arizona-I filter. This filter, when convolved with the response curve of the Mont4k CCD, yields a transmission range very similar to the Cousins-I band.

WASP-10 is a K5 dwarf at a distance of $90 \pm 20$ pc with an apparent V magnitude of $V=12.7$ (Christian et al. 2008). The conditions during observation were nearly photometric, with some high cirrus clouds, and our target field was far from the contaminating light of the moon. We utilized exposure times of just 10 seconds, with a read out time of approximately 20 seconds for a total sampling period of approximately 30 seconds. Seeing ranged from 1$^{\prime\prime}$.1 to 1$^{\prime\prime}$.5, which is typical for this site. Due to excellent autoguiding, there was less than 5 pixels (2.15\arcsec) of image wander over the course of the night. The airmass of our observations ranged between 1.30 at the start of our observations to 1.02 four hours later at the end of our observations. The onboard clock is synched to GPS every few seconds in order to ensure accurate absolute time keeping.

\begin{figure}[htp]
\centering
\includegraphics[scale=0.65]{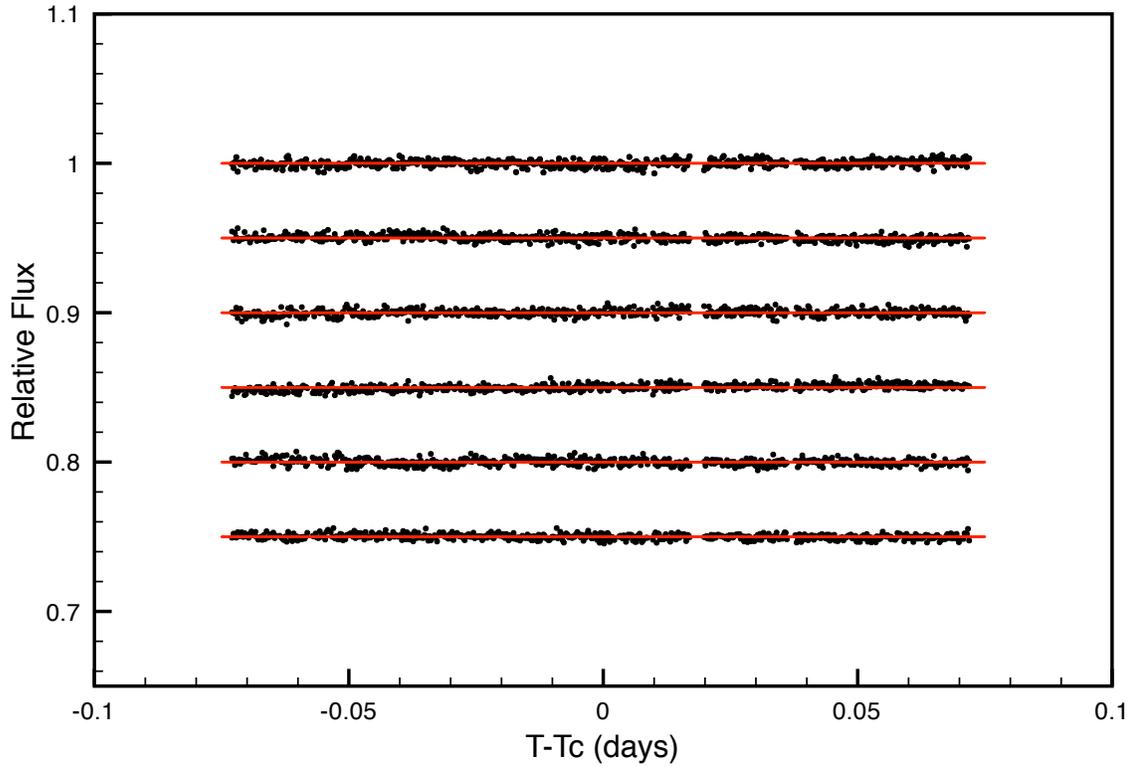}
\caption{Plot of the reference stars used to normalize the transit as a function of time. RMS scatter of these stars is approximately 2.0 mmag, which is typical for this instrumental setup and reduction. The flux of each reference star was normalized by the average fluxes of the other three reference stars. The two small ($\approx 2$ min) gaps in the data are due to our instrument script stopping and then restarting, and not due to weather. 
}
\label{RefStars}
\end{figure}

Each of the 579 images were bias-subtracted, flat-fielded, and bad pixel-cleaned in the usual manner. Aperture photometry was performed using the aperture photometry task PHOT in the IRAF DAOPHOT package.\footnote{IRAF is distributed by the National Optical Astronomy Observatories, which are operated by the Association of Universities for Research in Astronomy, Inc., under cooperative agreement with the National Science Foundation.} A 4$\arcsec$.3 aperture radius (corresponding to 10.0 pixels-see light curve shown in figure 1) was adopted because it produced the lowest scatter in the resultant lightcurve. Several combinations of reference stars were considered, but six were selected for the final reduction because of the low RMS ($\sim 2.0$ mmag) scatter. These reference stars were all distributed a few arcminutes east of the target, but scattered both slightly North and slightly South of WASP-10. We were unable to find sufficiently bright reference stars to the west of WASP-10 without increasing our RMS scatter. We show the time series light curve for each reference star (normalized to the average flux of the other reference stars) in Figure \ref{RefStars}.

\begin{figure}[htp]
\centering
\includegraphics[scale=0.65]{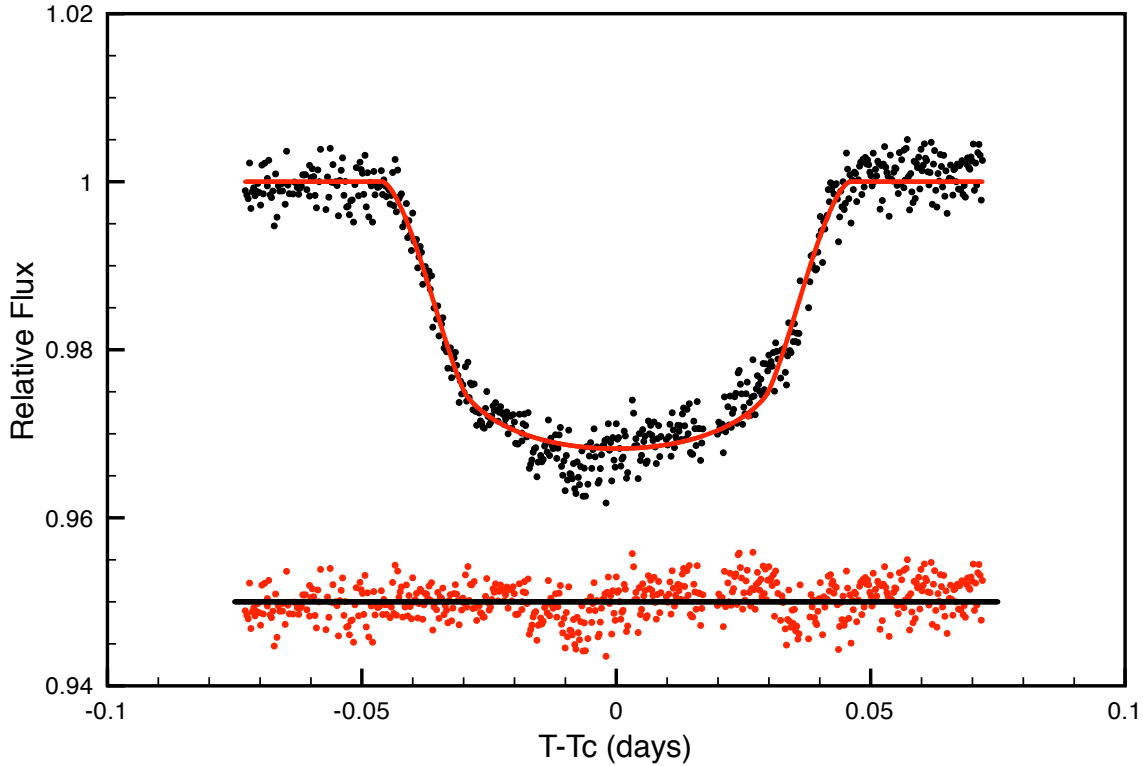}
\caption{Light curve of a transit of WASP-10b taken on 14 October 2009 at the University of Arizona's 1.55 meter Kuiper telescope on Mt Bigelow with the Mont4k CCD and the I filter. Standard deviation of the residuals (shown below the transit light curve) is approximately 2 mmag.  The transit was fit with the method of Mandel and Agol (2002), varying the central time of transit, planet to star radius ratio, and the impact parameter. Quadratic limb darkening coefficients for our model were taken from Claret (2000) for the I-band as 0.3678 and 0.2531. Our planet to star radius ratio of $0.16754 \pm 0.00060$ agrees with the results of Christian et al. (2008) and not the significantly smaller $0.15918^{+0.00050}_{-0.00115}$ result from Johnson et al. (2009). Our central transit time of HJD $2455118.66663 \pm 0.00060$ is in complete agreement with the linear ephemeris of Johnson et al. (2009).  }
\label{transit}
\end{figure}

We applied no sigma clipping rejection to the reference stars or WASP-10; all datapoints were used in the analysis. The final light curve for WASP-10 was normalized by division of the weighted average of the four reference stars. The unbinned residual time series in Figure \ref{transit} has a photometric RMS of 2.0 mmag, which is typical for the Mont4k on the 61-inch (Kuiper) telescope for our high S/N transit photometry pipeline (Dittmann et al. 2009a and b, Scuderi et al. (2010)). 

\section{Analysis}
We fit the transit light curve with the method prescribed by Mandel and Agol (2002), fitting for impact parameter, central time of transit, and planet to star radius ratio. The fit and residuals from the fit are shown in Figure \ref{transit}. Linear and quadratic limb darkening coefficients in the I band were taken from Claret (2000) as 0.3678 and 0.2531 respectively. In order to estimate the errors for our fits, we generated and fit 1000 fake data sets by taking our measured data points and adding white noise with a standard deviation equal to the instrumental standard deviation ($\approx 0.002$ mag) of our data points. The results of our fit, and those of Christian et al (2008) and Johnson et al. (2009) are shown in Table \ref{results}. 

\begin{figure}[htp]
\centering
\includegraphics[scale=0.65]{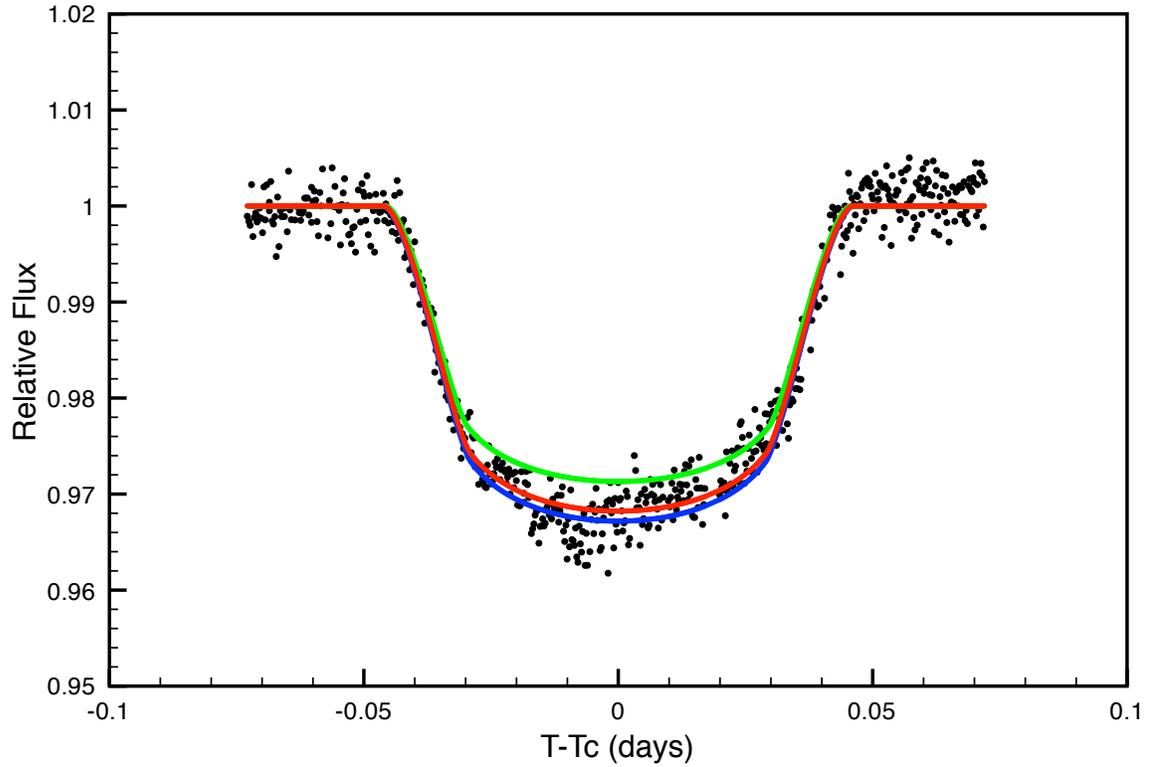}
\caption{Here, we overplot our best fit model (red), with the model of Christian et al. (2008) in blue and Johnson et al. (2009) in green. Our model, while falling between those of Christian et al. (2008) and Johnson et al. (2009) is significantly closer to Christian et al. (2008).
}
\label{comparison}
\end{figure}

The radius measurements of Christian et al. (2008) and Johnson et al. (2009) are in significant ($5\sigma$) disagreement. We find that our measurement, while between these previous results, is significantly closer to the result of Christian et al. (2008) than that of Johnson et al. (2009). We show a comparison between the different models in Figure \ref{comparison}. We find a planet to star radius ratio of $0.16754 \pm 0.00060$, consistent with the result of Christian et al. (2008) of  $0.17029 \pm 0.002$ but not with Johnson et al. (2009)'s of $0.15918^{+0.00050}_{-0.00115}$. However, we note that both our data and Christian et al. (2008)'s data has higher RMS scatter than Johnson et al. (2008)'s 0.5 mmag scatter data set. Furthermore, we find it unlikely that Johnson et al. (2009)'s data is contaminated by nearby stars as the brightest star within 40" of WASP-10 is over 100x fainter than WASP-10. However, if our result and those of Christian et al. (2008) are correct, then it is clear that WASP-10b is significantly inflated beyond expectations from theoretical modeling. This, combined with its nonzero eccentricity, make the WASP-10 system an interesting system and in need of further follow-up observations.

\section{Discussion}
\begin{figure}[htp]
\centering
\includegraphics[scale=0.65]{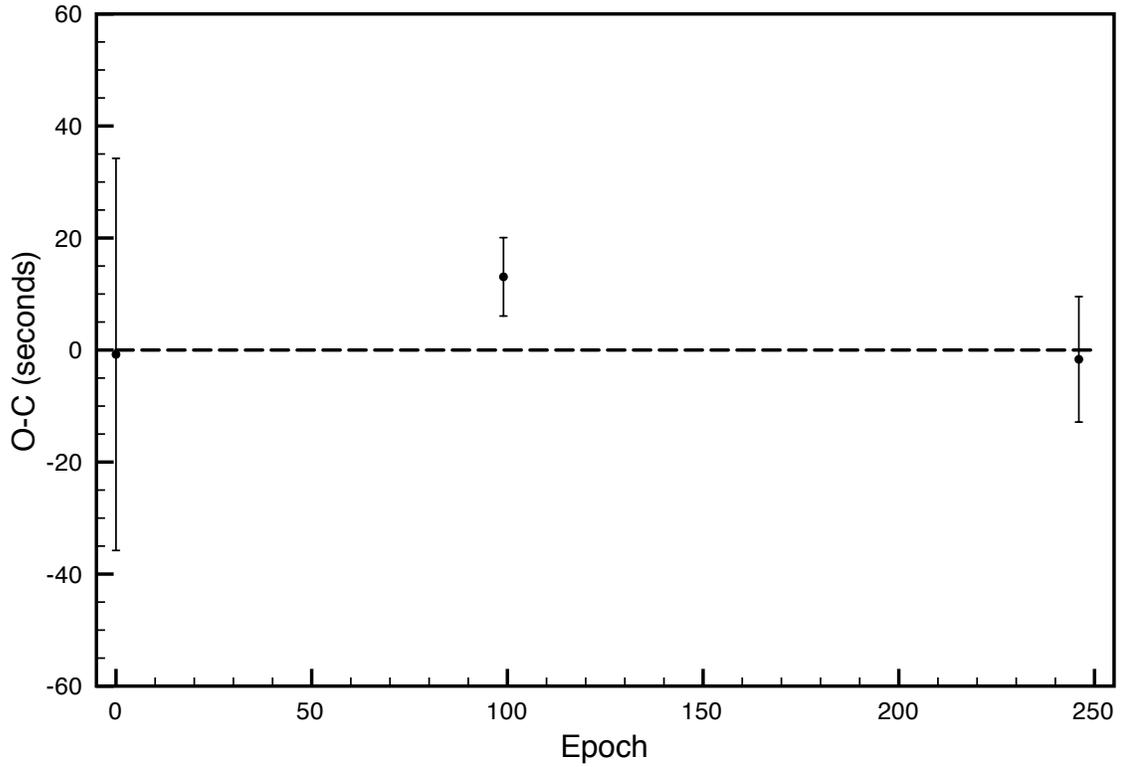}
\caption{O-C plot of the transit times of Christian et al. (2008), Johnson et al. (2010), and our transit point. We use our best fit ephemeris of $T_c = 2454357.858089+ E*3.092717$. We find no evidence for significant deviations from a linear ephemeris over long (246 orbit) timescales. 
}
\label{OC}
\end{figure}

Since WASP-10b has a significantly nonzero eccentricity orbit despite being older than its circularization timescale (Jackson et al. 2008), then it is possible that a third body is present in the system acting as an eccentricity pump. This body would perturb the orbit of WASP-10b, and could lead to measurable transit timing variations (TTVs), as theoretically investigated by Haghighipour et al. (2008). By using our the transit times of Christian et al. (2008), Johnson et al. (2010), and our data point, we are able to investigate a time scale of 246 orbits, or $\approx 760$ days. We fit a line to the available transit times, and find a refined best fit period of $3.092717 \pm 0.000007$ days, only $2 \sigma$ lower than that found by Christian et al. (2008). Figure \ref{OC} shows an observed-calculated (OC) plot for all currently published transit points, using our linear model. We find no large scale deviations ($> 20$s) from a linear ephemeris and therefore conclude that there are not likely to be large scale TTVs in the WASP-10 system and there is also unlikely to be a large perturber acting as an eccentricity pump. However, we note that TTVs require many observations to definitively rule out, as individual variations can periodically be small.


\section{Conclusions}
We have investigated one follow-up transit of WASP-10b in order to investigate the ($5\sigma$) discrepancy between the planet radii measurements of Christian et al. (2008) and Johnson et al. (2009). We have found that, while Johnson et al. (2009) apparently collected more precise data, our radii measurement agrees with that of Christian et al. (2008) and that WASP-10b does appear to be inflated beyond the level expected from theoretical models. Furthermore, we have no evidence that WASP-10b's eccentricity is due to perturbation by a third body. If there was a third body in the system, then we would expect successive transits of WASP-10b to exhibit signs of transit timing variations. However, we have found that our ephemeris is consistent to within $2 \sigma$ of Christian et al. (2008) and Johnson et al. (2009)'s ephemeris, and therefore large TTVs are unlikely in this system.

\acknowledgments

JD and LMC are supported by a NSF Career award and the NASA Origins program.

{\it Facilities:} \facility{Kuiper 1.55m}. \\
\\ {\bf References}\\\
\bibliography
BBakos, G. \'A., Torres, G. P\'al, A., et al., 2010, ApJ, 710, 1724 \\
Ballard, S. Christiansen, J.L, Charbonneau, D., et al. 2010, AAS, 425 \\
Baraffe, I., Chabrier, G., and Barman, T. 2008, A\&A, 482, 315 \\
Borucki, W., Koch, D., Basri, G., et al., 2008, Exoplanets: Detection, Formation and Dynamics, Proceedings of the International Astronomical Union, IAU Symposium, Vol. 249, 17\\
Burrows, A., Hubeny, I., Budaj, J., \& Hubbard, W. B. 2007, ApJ, 661, 502 \\
Butler, R.P., Vogt, S.S., Marcy, G.W., et al. 2004, ApJ, 617, 580 \\
Christian, D.J., Gibson, N.P., Simpson, E.K., et al., 2008, MNRAS, 392, 1585 \\
Claret, A., 2000, A\&A, 363, 1081 \\
Deming, D. Harrington, J., Laughlin, G., et al., 2007, ApJ, 667, L199 \\
Dittmann, J.A., Close, L.M., Green, E.M., Scuderi, L.J., Males, J.R. 2009a, ApJ, 699, L48.\\
Dittmann, J. A., Close, L.M., Green, E.M., Fenwick, M. 2009b, ApJ, 701, 756 \\
Fortney, J.J., Marley, M.S., and Barnes, J.W., 2007, ApJ, 659, 1661 \\
Jacokson, B., Greenberg, R., Barnes, R. 2008, ApJ, 678, 1396 \\
Johnson, J.A., Winn, J.N., Cabrera, N.E. et al., 2009, ApJ, 692, L100 \\
Johnson, J.A., Winn, J.N., Cabrera, N.E., et al., 2010 ApJ, 711, L1 \\
Mandel, K., \& Agol, E. 2002, ApJ, 580, L171\\
Maness, H.L, Marcy, G.W., Ford, E.B., et al., 2007, Pub. Ast. Soc. Pac., 119, 90 \\
Pont, F., Gilliland, R.L, Knutson, H., et al. 2009, MNRAS, 393, L6 \\
Scuderi, L.J., Dittmann, J.A., Close, L.M, et al. 2009, ApJ, in press (arXiv:0907.1686) \\
\clearpage

\begin{table}
\begin{center}
\caption{Parameters of the WASP-10 system}\label{results}

\begin{tabular}{crrr}
\tableline\tableline
  Parameter [units] & Value & Reference \\
  \tableline
 $T_c$ (HJD-2,400,000) & $54357.85808^{+0.00041}_{-0.00036}$ & Christian et al. (2008) \\
 $T_c$ (HJD-2,400,000) & $54664.037295 \pm 0.000041$ & Johnson et al. (2009) \\
 $T_c$ (HJD-2,400,000) & $55118.66663 \pm 0.00013$ & This work \\
 $R_p/R_*$ & $0.17029 \pm 0.002$ & Christian et al. (2008) \\
 $R_p/R_*$ & $0.15918^{+0.00050}_{-0.00115}$ & Johnson et al. (2009) \\
 $R_p/R_*$ & $0.16754 \pm 0.00060$ & This work \\
 $P$ (days) & $3.0927636^{+0.0000094}_{-0.000021}$ & Christian et al. (2008) \\
 $P$ (days) & $3.0927616 \pm 0.0000056$ & Johnson et al. (2009) \\
 $P$ (days) & $3.092717 \pm 0.000007$ & This work \\
\tableline
\label{results}
\end{tabular}


\end{center}
\end{table}

\clearpage

\clearpage

\clearpage

\clearpage

\clearpage

\end{document}